\documentclass[aps, prb, reprint, superscriptaddress, amsmath, amssymb, floatfix, twocolumn]{revtex4-1}
\bibpunct{[}{]}{;}{n}{}{}

\usepackage{xcolor}
\usepackage{graphicx}
\usepackage{dcolumn}
\usepackage{bm}
\usepackage{courier}
\usepackage{pgfplotstable}
\pgfplotsset{compat=1.17}
\usepackage{float}

\begin{document}

\title{Two-dimensional ferroelectrics from high throughput computational screening}

\author{Mads Kruse}
\affiliation{CAMD, Department of Physics, Technical University of Denmark, DK-2800 Kongens Lyngby, Denmark}%
\author{Urko Petralanda}
\affiliation{CAMD, Department of Physics, Technical University of Denmark, DK-2800 Kongens Lyngby, Denmark}%
\author{Morten N. Gjerding}
\affiliation{CAMD, Department of Physics, Technical University of Denmark, DK-2800 Kongens Lyngby, Denmark}%
\author{Karsten W. Jacobsen}
\affiliation{CAMD, Department of Physics, Technical University of Denmark, DK-2800 Kongens Lyngby, Denmark}%
\author{Kristian S. Thygesen}
\affiliation{CAMD, Department of Physics, Technical University of Denmark, DK-2800 Kongens Lyngby, Denmark}%
\author{Thomas Olsen}
\email{tolsen@fysik.dtu.dk}
\affiliation{CAMD, Department of Physics, Technical University of Denmark, DK-2800 Kongens Lyngby, Denmark}%

\date{\today}

\begin{abstract}
We report a high throughput computational search for two-dimensional ferroelectric materials. The starting point is 252 pyroelectric materials from the computational 2D materials database (C2DB) and from these we identify 64 ferroelectric materials by explicitly constructing adiabatic paths connecting states of reversed polarization. In particular we find 49 materials with in-plane polarization, 8 materials with out-of-plane polarization and 6 materials with coupled in-plane and out-of-plane polarization. Most of the known 2D ferroelectrics are recovered by the screening and the far majority of the new predicted ferroelectrics are known as bulk van der Waals bonded compounds, which implies that these could be experimentally accessible by direct exfoliation. For roughly 25{\%} of the materials we find a metastable state in the non-polar structure, which could have important consequences for the thermodynamical properties and may imply a first order transition to the polar phase. Finally, we list the magnetic pyroelectrics extracted from the C2DB and focus on the case of VAgP$_2$Se$_6$, which exhibits a three-state switchable polarization vector that is strongly coupled to the magnetic excitation spectrum.
\end{abstract}

\maketitle
\section{Introduction}
Ferroelectric materials are characterized by having a spontaneous electric polarization that is switchable by means of an external field \cite{ferroBook}. This property makes them suitable for non-volatile memory applications, which may be based on ferroelectric tunnel junctions, ferroelectric random access memory or ferroelectric field effect transistors \cite{ferrocapacitors,memorydevice,tunneljunctions,Mikolajick2021}. While the large dielectric constant of ferroelectrics makes them ideal constituents of standard capacitors, it has also been demonstrated that ferroelectrics can acquire a negative capacitance under particular circumstances and this property may be utilized to reduce the energy consumption of conventional electronics significantly \cite{Khan2015,Zubko2016,Hoffmann2021}. Finally, the inherent pyroelectric and piezoelectric properties of ferroelectrics make them useful for a wide range of sensor and actuator applications \cite{ferrocapacitors,ferroactuartorsandsensor}. The recent discovery of monolayer ferroelectrics \cite{SnTe_experiment, SnS_experiment, SnSe_experiment, SnSedomainswitching} has initiated fundamental interest in the basic properties of 2D ferroelectricity and spurred hope that devices based on ferroelectrics may undergo a dramatic size reduction. There is currently an intense effort to discover new 2D ferroelectrics with optimized properties and the list of materials that have been characterized experimentally includes 
In$_{2}$Se$_{3}$ \cite{In2Se3experiment, alphaIn2Se3experiment}, CuInP$_{2}$Se$_{6}$ \cite{CIPS_experiment, CIPS_nature}, MoTe$_2$ \cite{Yuan2019},  NiI$_{2}$ \cite{Song2022}, SnTe \cite{SnTe_experiment}, SnSe \cite{SnSe_experiment,SnSedomainswitching} and SnS \cite{SnS_experiment}.

Devoid of the challenges that the synthesis of new 2D materials usually present, density functional theory (DFT) has emerged as an important tool to predict new ferroelectrics and characterize their properties. The compounds studied by DFT include in-plane ferroelectrics such as the group-IV monochalcogenides GeSe, GeS, SnS, SnSe, GeTe, SiTe, SnTe \cite{polarization_2D_monochalcogenides, XTes_paper}, the niobium oxyhalides (NbOI$_{2}$, NbOCr$_{2}$, NbOBr$_{2}$)\cite{Niobiumoxides},
the magnetic vanadium oxyhalides VOX$_{2}$ (X = I, Cl, Br, F)\cite{VOX2_multiferroics, VOF2paper} as well as ClGaTe and $\gamma$-SbX (X = As, P)\cite{ClGaTe_paper, gammaSbX}. In addition, the transition metal phosphorus chalcogenides ABP$_{2}$X$_{6}$ (A = Cu/Ag, B = In/Bi/Cr/V, X = S/Se)\cite{CIPS_nature, ABP2X6_paper1,CrCuS2P6, ABP2X6_magnetic} have been predicted to exhibit out-of-plane polarization and
thus overcome the fundamental thickness limit below which out-of-plane polarization vanishes in standard perovskite thin films \cite{Junquera2003}. 
Other compounds, such as $\alpha$-In$_{2}$S$_{3}$ and related III$_2$-IV$_3$ compounds have been predicted to have both in-plane and out-of-plane components of polarization, which are strongly intercorrelated \cite{alphaIn2Se3_theory}. Finally, the strong confinement of electrons in 2D materials makes screening of out-of-plane polarization inefficient and have led to the prediction of the ferroelectric metals CrN and CrB \cite{Luo2017}.

For three-dimensional compounds there has been a vivid search for new ferroelectrics with optimal properties. For example with respect to lead-free energy storage materials \cite{Zhang2020} or low switching barriers for memory applications \cite{Shi2022}. To this end, high throughput DFT calculations has proven a powerful strategy that can be used to rapidly screen thousands of materials for desired properties. For the case of ferroelectrics, however, it is not always straightforward to identify good candidate materials in a high throughput framework, since one has to demonstrate the existence of an {\it a priori} unknown switching path. In Ref. \cite{Garrity2018}, 16 ferroelectrics were identified from 2750 polar materials taken from the Inorganic Crystal Structure Database (ICSD). More recently, a high throughput project based on 67,000 materials from the Materials Project database \cite{Smidt2020} identified 126 new ferroelectrics. The primary success criterion of such search strategies is the subsequent experimental demonstration of ferroelectric properties of the predicted materials. Such validation may pose major experimental challenges and in some cases predicted ferroelectrics have been shown to condense into non-polar phases during synthetization \cite{Acharya2020}. Regarding 2D materials, there seems to be a lack of systematic high throughput search for new ferroelectrics. One exception is a recent study where 60 new 2D ferroelectrics were identified from lattice decoration of a certain prototype \cite{Ma2021}, but it is again not completely clear whether such an approach actually leads to experimentally realizable materials. In the present work, we present a high throughput screening for 2D ferroelectrics based on the C2DB, which contains materials obtained from lattice decoration as well as materials that are expected to be easily exfoliable from bulk van der Waals bonded materials. In addition to well known 2D ferroelectrics we find a wide range of new 2D ferroelectric materials with both in-plane, out-of-plane and mixed polarization. The far majority of the materials are known experimentally in their 3D bulk form and it is reasonable to expect that several of these may be validated experimentally in the future.

The paper is organized as follows. In Sec. \ref{sec:theory} we briefly outline the theoretical background and state the computational details applied in the present work. In Sec. \ref{sec:results} we explain the workflow used to obtain ferroelectric materials from C2DB and discuss the results. In particular, we provide a detailed description of the adiabatic path required to assign a polar insulator as ferroelectric and classify materials according to direction of polarization with respect to the atomic plane and the stability of phonons in the non-polar reference state. Finally, in Sec. \ref{sec:mag} we list the magnetic polar materials in C2DB and discuss the ferromagnetic and ferroelectric compound VAgP$_2$Se$_6$, which was not found by our workflow. Sec. \ref{sec:out} contains an outlook and conclusion.

\section{Theoretical background}\label{sec:theory}
\subsection{Formal Polarization}
The defining property of ferroelectric materials is the presence of a spontaneous electrical polarization, which is switchable by application of an external electric field. While the bulk polarization is not well defined in bulk materials one may define the spontaneous polarization as the change in polarization with respect to a non-polar state that can be adiabatically connected to the polar phase. According to the modern theory of electrical polarization \cite{King-Smith1993,Vanderbilt1993,Resta2007}, one can calculate the formal 2D polarization for a system in a given configuration as
\begin{align}
\mathbf{P} =e\bigg(\frac{1}{A}\sum_aZ_a\mathbf{r}_a-\sum_n\int_\mathrm{BZ}\frac{d\mathbf{k}}{(2\pi)^2}\langle u_{n\mathbf{k}}|i\nabla_\mathbf{k}|u_{n\mathbf{k}}\rangle\bigg) 
\label{eq:formal_polarization}
\end{align} 
where $e$ denotes the electron charge, $A$ is the unit cell area, $Z_a$ is the charge of nucleus $a$ (including core electrons), $u_{n\mathbf{k}}$ are Bloch states represented in a smooth gauge such that the $k$-space derivative is well defined and the sum runs over occupied states. 
The formal polarization is only defined modulo $e\mathbf{R}_i/A$, where $\mathbf{R}_i$ is an arbitrary lattice vector. This is due to the fact that the nuclei positions can be chosen in any unit cell and the Brillouin zone integral may change by $\mathbf{R}_i/A$ by a gauge transformation of the Bloch states. Nevertheless, differences in polarization along any adiabatic path (continuous change of atomic positions where the system remains insulating) is well defined since one may track the polarization along a particular branch. One can therefore define the spontaneous polarization as 
\begin{align}
\mathbf{P}_\mathrm{s} = \int_{\lambda=0}^{\lambda=1}d\lambda\frac{d\mathbf{P^\lambda}}{d\lambda}.
\end{align}
where $\lambda$ parameterizes an adiabatic path between a non-polar reference structure ($\lambda=0$) and the polar ground state ($\lambda=1$).

\subsection{Coercive Field}
The coercive field is defined as the electric field required to switch a ferroelectric material between two different polarization states. In reality this is likely to involve complicated structural reorganization, which is typically dominated by migration of domain walls \cite{SnS_experiment,DW_DFT_Coercive2022,Merz1954,Merz1956,Miller1958}. In the present work we simply calculate the field required for coherent monodomain switching, which may be orders of magnitude larger than the actual field required for polarization switching mediated by domain wall motion. Nevertheless, the field provides a rough measure of the polarization stiffness in the material and yields an upper bound for the true coercive field. 
\section{Results}\label{sec:results}
\subsection{Symmetries}
\begin{figure*}[tb]
    \centering
    \includegraphics[width=1.9\columnwidth]{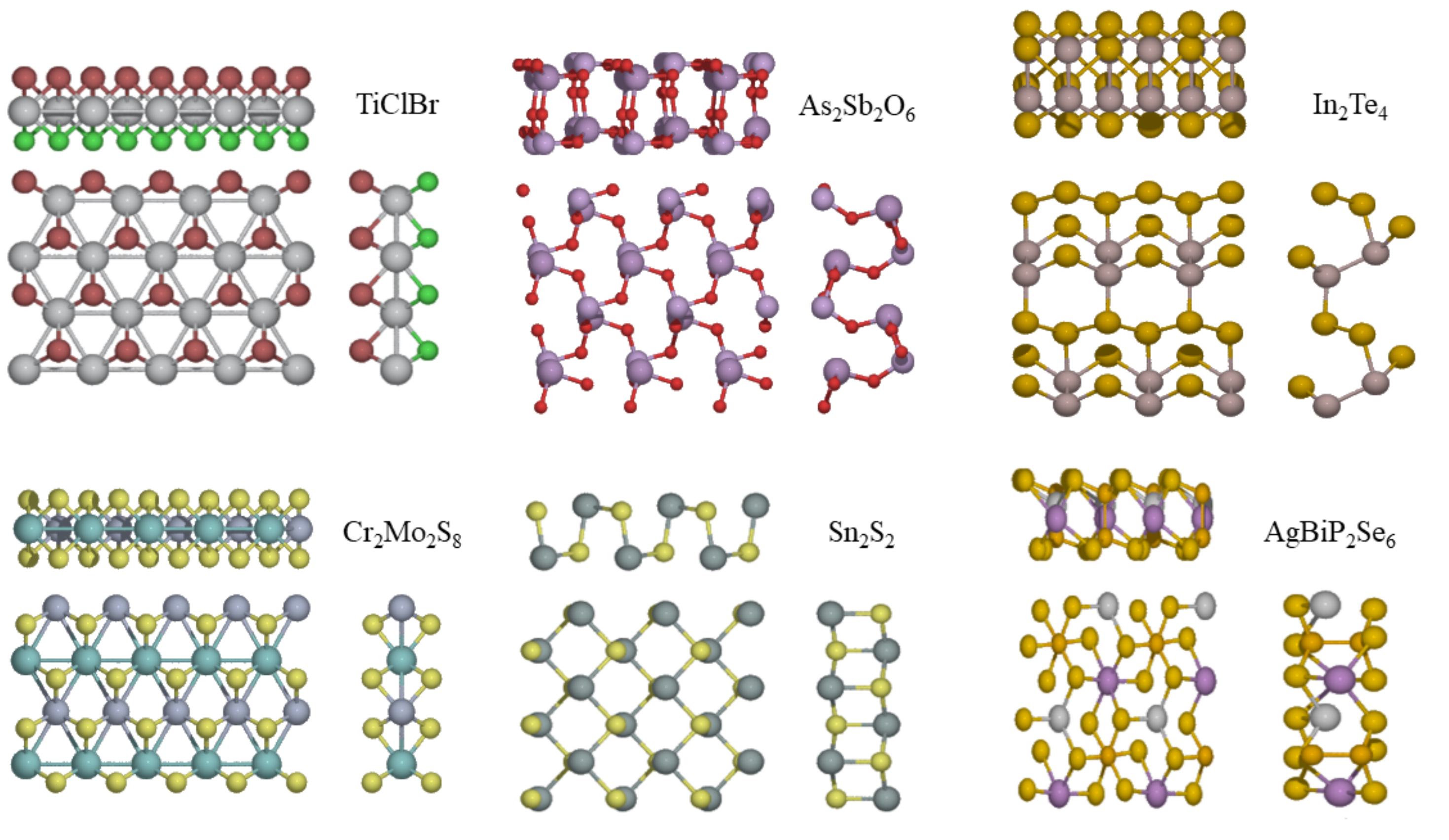}
    \caption{Prototypes of polar materials. The first column depicts the 2 non-ferroelectric pyroelectrics TiClBr and Cr$_2$Mo$_2$S$_8$. The second column depicts two in-plane ferroelectrics As$_2$Sb$_2$O$_6$ and Sn$_2$S$_2$. The third column depicts the ferroelectric In$_2$Te$_4$, which has polarization with both in-plane and out-of-plane components and AgBiP$_2$Se$_6$, which has purely out-of-plane polarization.}
    \label{fig:materialprototypes}
\end{figure*}

The coercive field for monodomain switching can be estimated by finding the minimum energy path connecting the two structures corresponding to $\lambda=0$ (non-polar) and $\lambda=1$ (polar), which may be obtained by the nudged elastic band method \cite{Henkelman2000}. In the presence of an electric field the energy will acquire a term proportional to $\boldsymbol{\mathcal{E}}\cdot\mathbf{P}$ and the the system will change polarization state when the force originating from the field matches the maximal slope of the energy along the path. If the field is applied parallel to the polarization the magnitude of the coercive field may then be calculated according to
\begin{equation}\label{eq:field}
\mathcal{E}_\mathrm{c} = \text{max}\left(\frac{dE(\mathbf{P}^\lambda)}{d|\mathbf{P}|}\right),
\end{equation}
where $E(\mathbf{P}^\lambda)$ is the energy per unit area along the path. We note that even in the realm of strict monodomain switching, this is an approximate approach since the true minimal path in the presence of an electric field may differ from that found without a field. 

\subsection{Computational details}
All the calculations in the present work were carried out using the electronic structure package \texttt{GPAW} \cite{Enkovaara2010,Larsen2017}, which applies the projector-augmented wave (PAW) method and a plane wave basis set. We have used the Perdew-Burke-Ernzerhof (PBE) functional, a plane wave cutoff of 800 eV, a k-point density of 12 {\AA} and a Fermi smearing of 0.05 eV. All structures were relaxed with a force tolerance of 1 meV{\AA}. For details on the implementation of Eq. \ref{eq:formal_polarization} we refer to Ref. \cite{Gjerding2021}.
\begin{figure*}[tb]
    \centering
    \includegraphics[width=1.9\columnwidth]{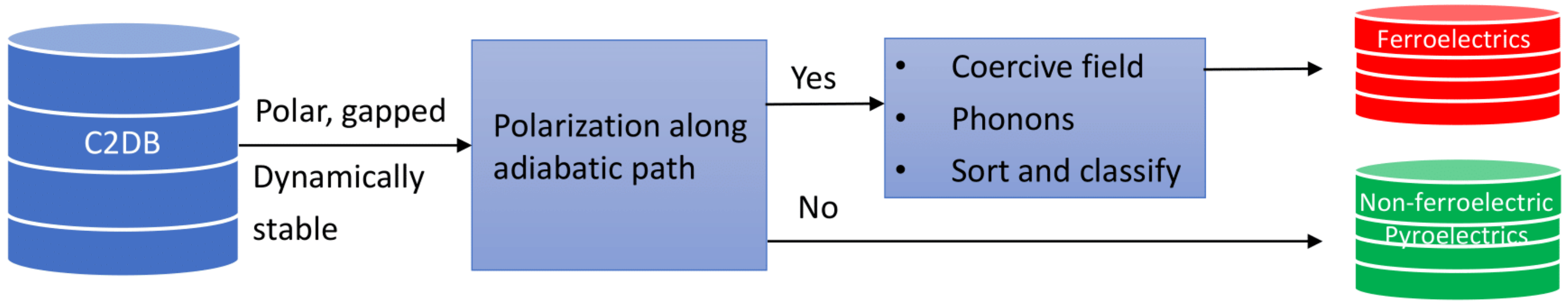}
    \caption{The figure summarises the workflow used in the study. We pick out all dynamically stable gapped materials with a polar point group. The materials for which an adiabatic polarization path exists are classified as ferroelectric and the ones where it is ill defined are classified as pyrolelectric. For ferroelectric materials additional computations are performed and the materials are further classified according to the direction of polarization (see main text).}
    \label{fig:workflow}
\end{figure*}

The starting point of the calculations performed in this work is all dynamically stable 2D insulators from the C2DB \cite{Haastrup2018} that have a polar point group. There are 252 materials satisfying these criteria, but many of these are not expected to be ferroelectric. For example, the Janus monolayers where a single transition metal atom is sandwiched between two {\it} different chalcogenides always have an out-of-plane dipole, which are not switchable by an external electric field. Such materials can obviously not be regarded as being ferroelectric, but one may easily calculate the spontaneous polarization, by simply integrating the electronic density weighted by a coordinate orthogonal to the layer over the entire system. For polar materials with an in-plane polar axis, this approach cannot be applied and we may only define the spontaneous polarization if an adiabatic path that connects the structure to a non-polar phase can be identified. In Fig. \ref{fig:materialprototypes} we show examples of various prototypical compounds in the C2DB exhibiting polar point groups.

For 2D ferroelectrics it is natural to distinguish between materials where the symmetries forbid components of the polarization in the plane, materials where the polarizations out of plane is forbidden and those where the polar axis is neither constrained to the in-plane or our-of-plane directions. For the ten crystallographic polar point groups $1$, $2$, $m$, $mm2$, $3$, $3m$, $4$, $4mm$, $6$ and $6mm$ only the first four may have an in-plane polar axis. In the cases of $2$ and $mm2$ the polarization is parallel to the two-fold axis, which may be either in-plane or out-of-plane. For $1$ there is no constraints on the polarization and for $m$ the polarization is confined to be in-plane if the mirror coincides with the atomic plane. As mentioned above, it is possible to calculate the spontaneous polarization for out-of-plane polar materials even if they are not ferroelectric, but we will not go into detail with these since this group is dominated by Janus monolayers, which have been studied elsewhere \cite{Riis-jensen2019}. In Fig. \ref{fig:workflow} we outline the workflow applied to classify the polar and gapped materials in C2DB.

\subsection{Definition of the adiabatic path}
\begin{figure}[tb]
    \centering
    \includegraphics[width=\columnwidth]{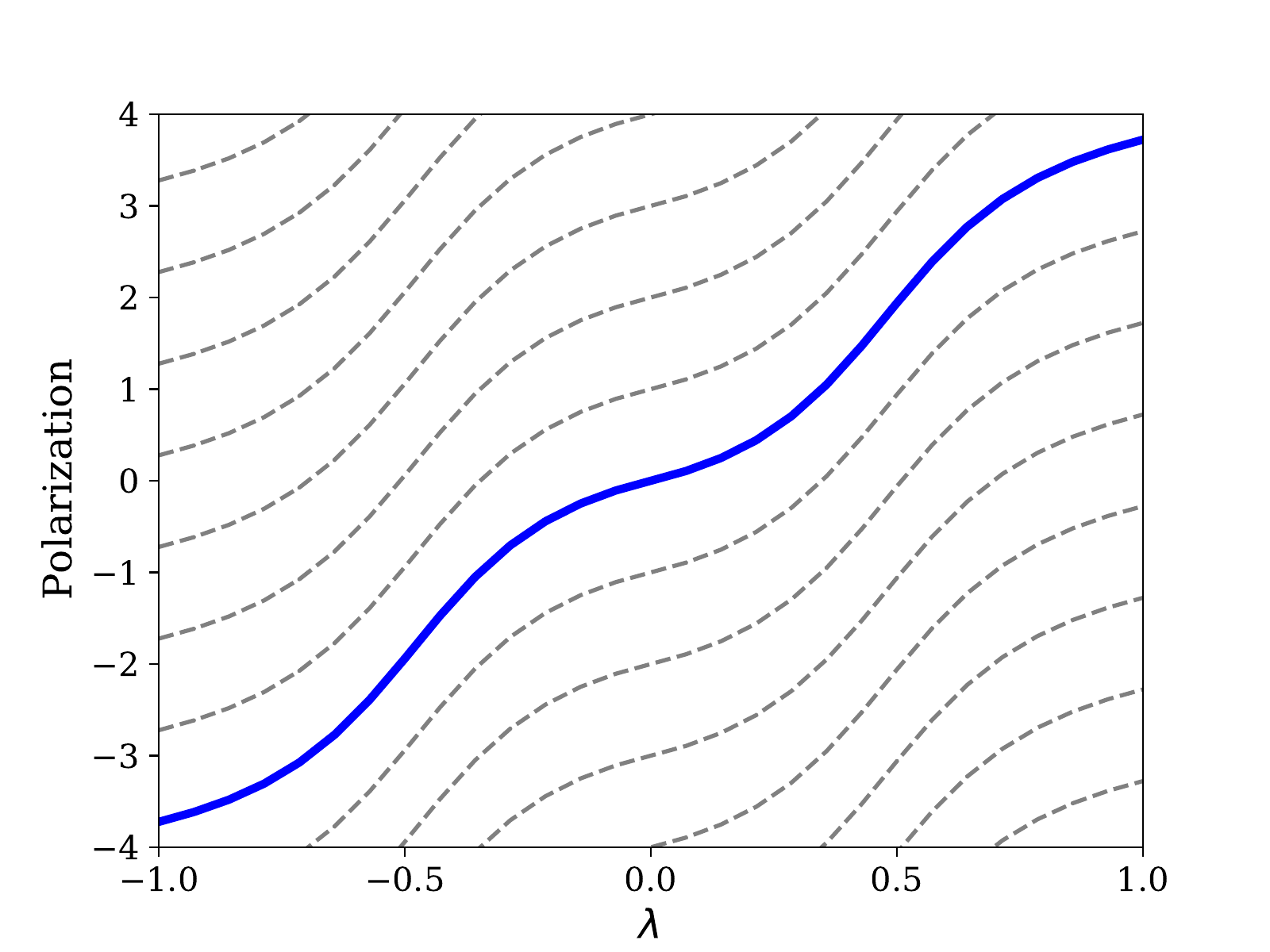}
    \caption{Polarization branches for As$_{2}$Se$_{3}$ in units of the lattice vector divided by unit cell area. For the adiabatic path we choose one branch (in this case the one that crosses zero polarization at $\lambda=0$) and the spontaneous polarization $P_\mathrm{s}$ is defined as the change in polarization between $\lambda=0$ and $\lambda=1$.}
    \label{fig:Polarization_shifted_As2Se3}
\end{figure}

For materials that do not have an out-of-plane polar axis one needs a non-polar reference state that can be adiabatically connected with the polar states in order to calculate the spontaneous polarization. We have used the module \texttt{evgraf} \cite{evgrafsoftware} to generate the  centrosymmetric structure of highest similarity to the polar structure \cite{Mahler2020} of all materials. This structure is then relaxed under the constraint of inversion symmetry and the result is taken as our reference ($\lambda=0$) non-polar structure. We then check if there is an adiabatic path connecting the two points by performing nudged elastic band (NEB) calculations \cite{Henkelman2000} connecting the polar and non-polar structures. In all calculations the unit cell of the polar structure is conserved. If the band gap remains finite along the path there is a well defined adiabatic path and we regard the material as being ferroelectric. We then proceed by performing single-shot DFT calculations for each structure on the path including the two endpoints and evaluate the formal polarization from the Berry phase formula \eqref{eq:formal_polarization} and for each point on the path we choose the polarization branch which deviates the least from the former point on the path. With this procedure the spontaneous polarization is calculated as the (branch fixed) change in polarization along the path. In Fig. \ref{fig:Polarization_shifted_As2Se3} we show an example of such a calculation for As$_2$Se$_3$, where the polarization is shown in units of the polarization quantum $e\mathbf{a}_1/A$ ($\mathbf{a}_1$ is the unit cell vector which is parallel to the polarization and $A$ is the unit cell area). This represents a rather extreme case where the spontaneous polarization is 3.5 times larger than the polarization quantum and highlights the importance of branch fixing on the adiabatic path. A naive single shot calculation of the polarization in the polar phase will yield an arbitrary point in the infinite polarization lattice. Moreover, in general the polarization is not guaranteed to vanish at $\lambda=0$, since non-polar structures may acquire a finite topological polarization with certain discrete values dictated by the symmetry of the polarization lattice \cite{Benalcazar2017a,Benalcazar2019}. We note that a linear interpolation between the polar and non-polar structures is sufficient for calculating the spontaneous polarization as long as the system remains insulating along the path. The NEB is therefore only required for extracting energy barriers between the non-polar and polar structures and for the calculation of the coercive field. 

\subsection{Ferroelectrics in C2DB}
From the 252 stable and gapped polar materials in C2DB we have identified 63 materials with a well-defined adiabatic path connecting the polar structure to a non-polar structure. The remaining  189 compounds are classified as non-ferroelectric pyroelectrics. Except for the magnetic materials, which are discussed below, the non-ferroelectric pyroelectrics will not be analyzed further in the present work. We will just mention that 63 of these have an in-plane polar axis, 119 have an out-of-plane polar axis and 7 have a polar axis that is neither constrained to the atomic plane or the direction orthogonal to the atomic plane. 
\begin{figure*}[tb]
    \centering
    \includegraphics[width=0.68\columnwidth]{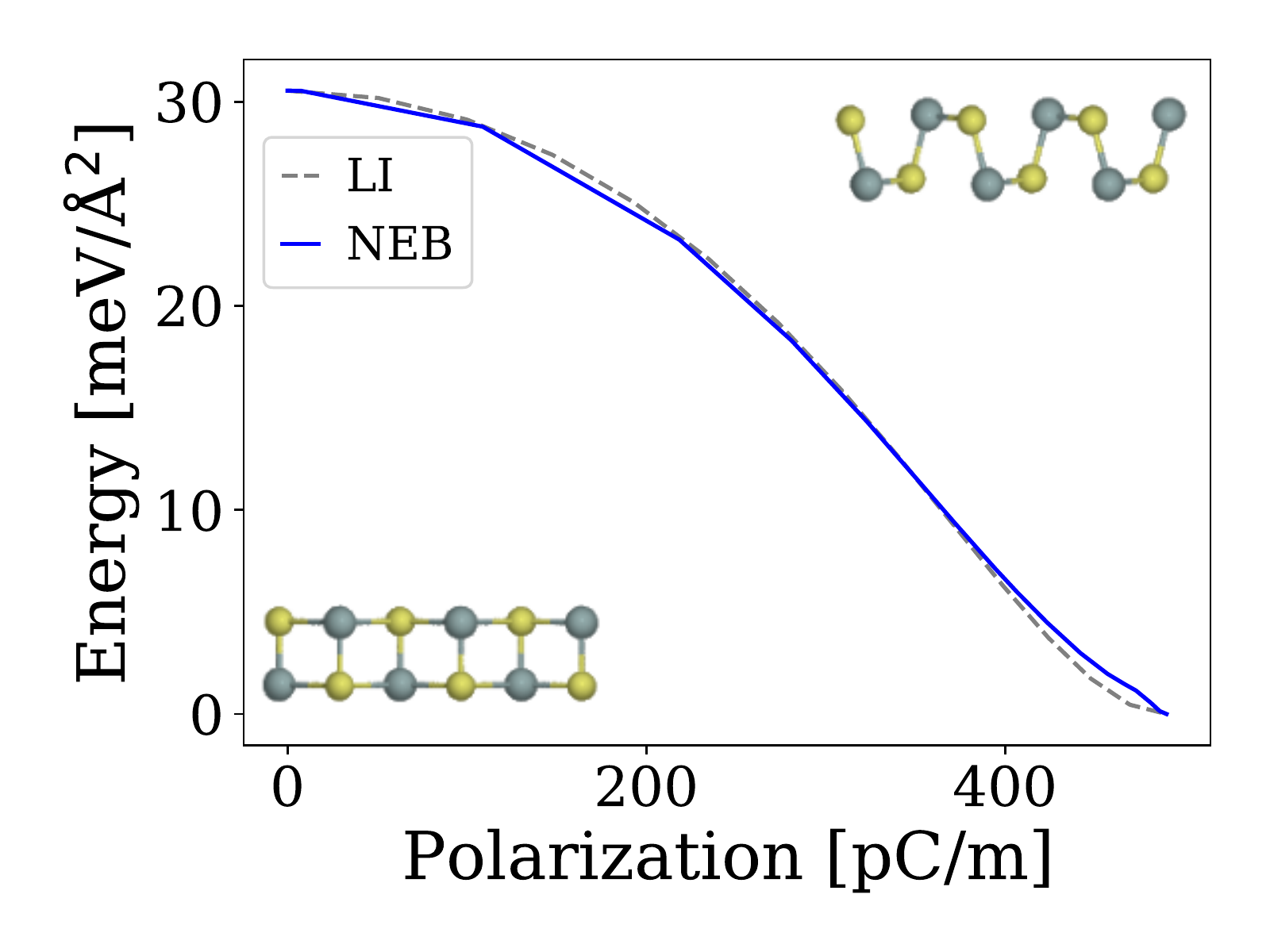}
    \includegraphics[width=0.68\columnwidth]{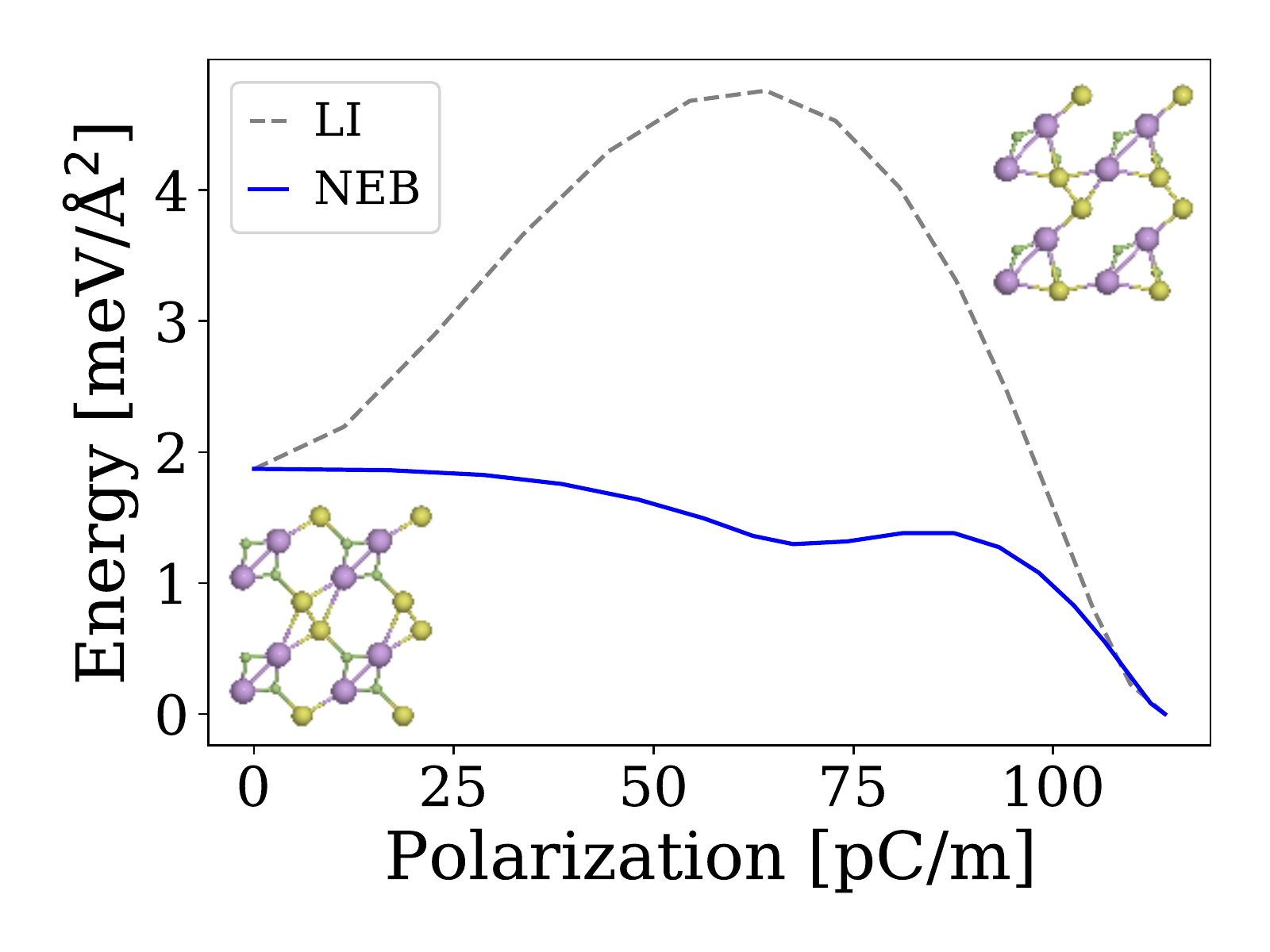}    
    \includegraphics[width=0.68\columnwidth]{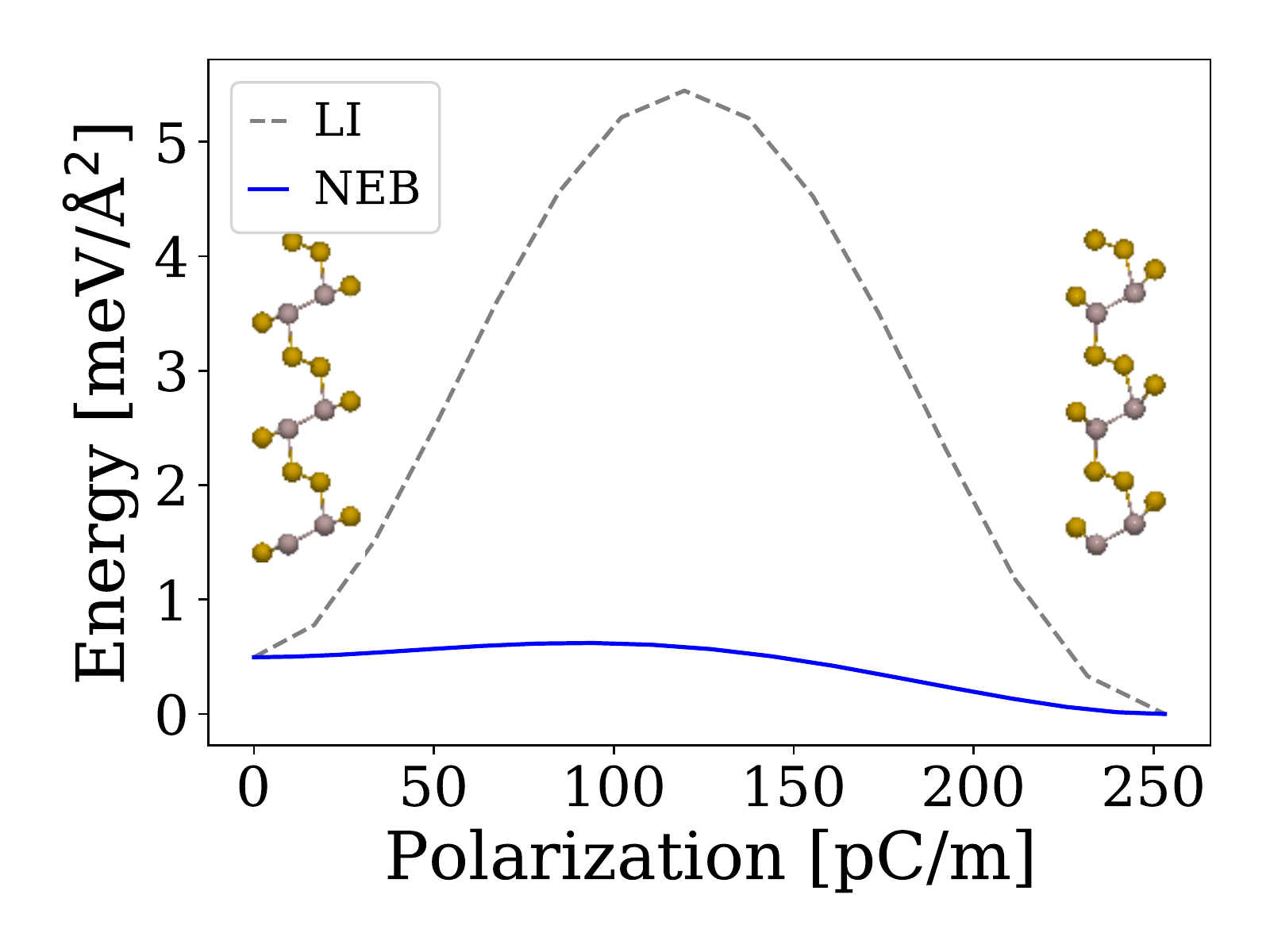}
    \caption{Energy versus polarization for three representative ferroelectrics. Left: Ge$_{2}$S$_{2}$ (in-plane polarization). Middle: F$_{2}$Li$_{2}$S$_{2}$ (in-plane polarization). Right: In$_{2}$Te$_{4}$, which has both in-plane and out-of-plane components of the spontaneous polarization. In all three cases we show the energy along the linearly interpolated path (LI) and energy along the path optimized by the nudged elastic band method (NEB). The insets show the side views of polar and non-polar states for Ge$_2$S$_2$ and In$_{2}$Te$_{4}$ and top views for F$_{2}$Li$_{2}$S$_{2}$.}
    \label{fig:E_vs_P}
\end{figure*}

There are two main reasons why one may fail to define a proper adiabatic path in the remaining materials. 1) Some point along the path becomes metallic. 2) The calculation breaks down because the centrosymmetric structure is unphysical with atoms situated on top of each other (for example in the case of Janus monolayers). We cannot exclude the existence of an adiabatic path in either of these cases since we always consider the non-polar phase to be centrosymmetric, which does not need to be the case. For example, for the Janus layers one may define a path from $P$ to $-P$ that does not involve a centrosymmetric structure. However, even in the unlikely situation where no gapless structure emerges along such a path, the energy barrier for the transition would be rather large and is not likely to be switchable under realistic conditions. Nevertheless, there may be other materials where we miss a relevant adiabatic path between $P$ and $-P$ that does not involve a centrosymmetric state, but we leave a systematic study of such cases to future work. It may also happen that the path passing through a centrosymmetric point is adiabatic and may be used to extract the spontaneous polarization, but the path is not necessarily the minimum energy path. This situation is realized in most of the out-of-plane ferroelectrics as will be discussed below.

\begin{figure}[b]
    \centering
    \includegraphics[width=\columnwidth]{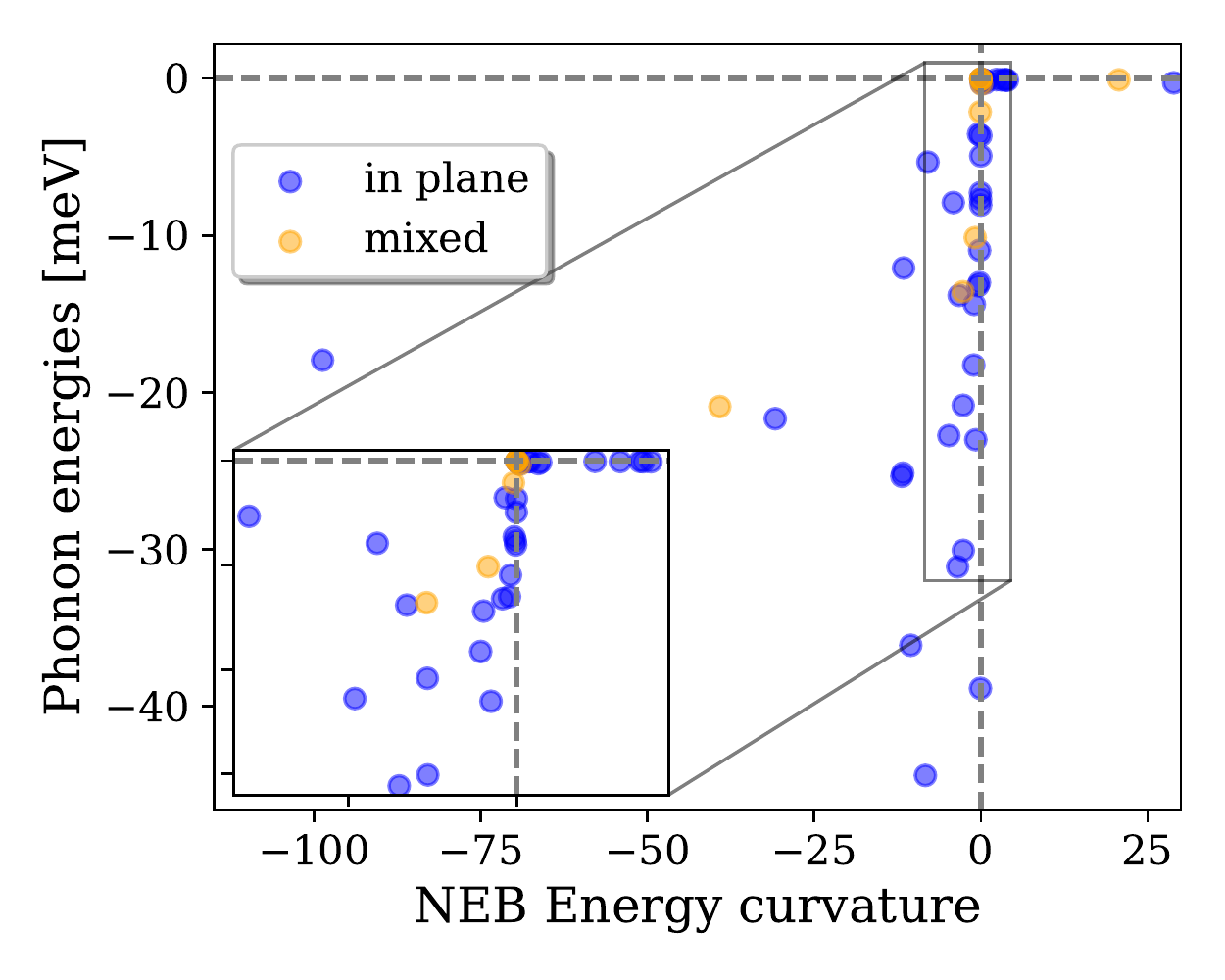}
    \caption{Lowest phonon frequency (imaginary modes represented as negative numbers) versus energy curvature of the NEB energy potential surface calculated in the non-polar phase. The points are color coded according to the direction of polarization with respect to the atomic plane.}
    \label{fig:Phonon_scatterplot}
\end{figure}
\begin{figure}[b]
    \centering
    \includegraphics[width=\columnwidth]{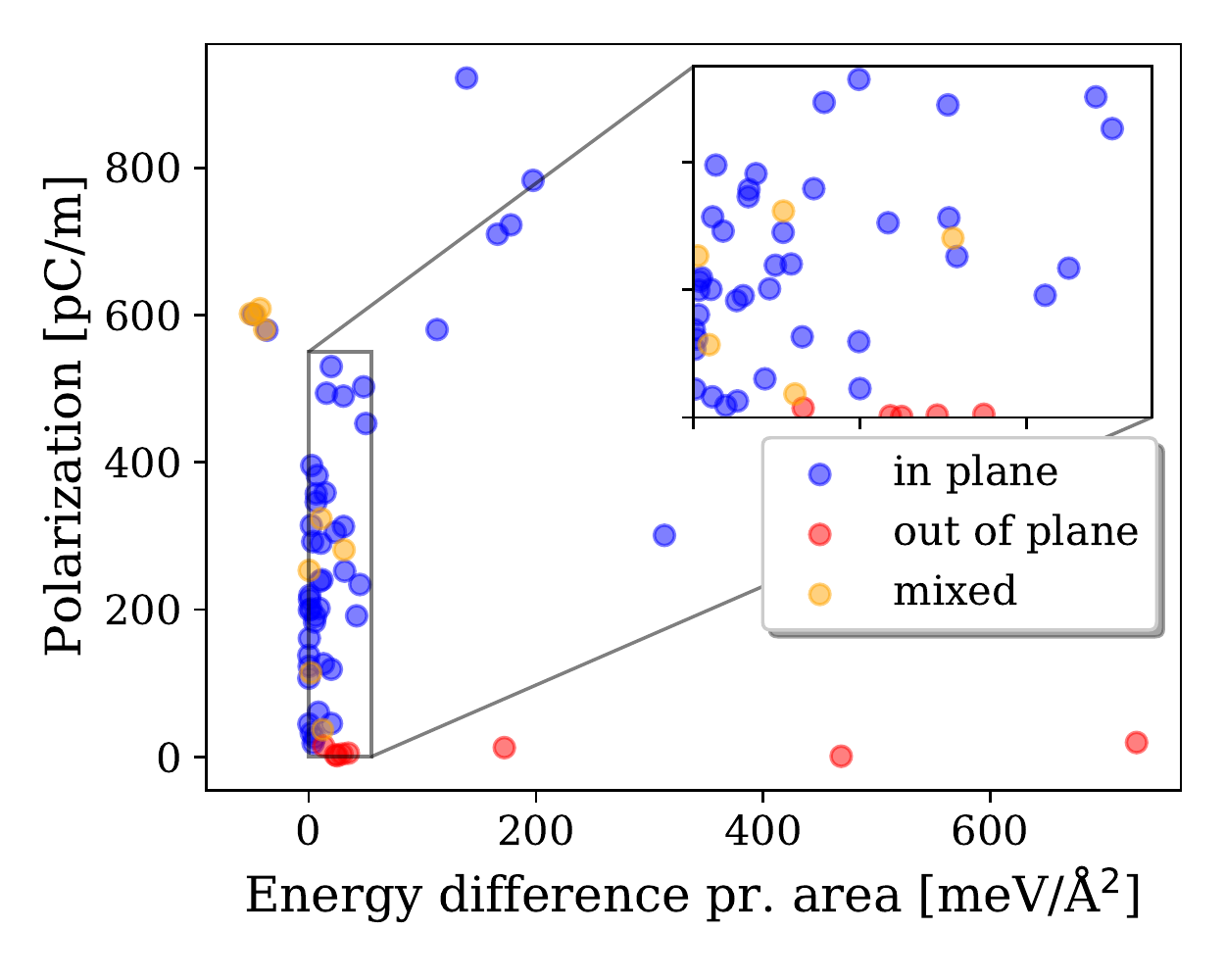}
    \caption{Magnitude of polarization versus the energy difference between the polar and non-polar phases. The points are color coded according to the direction of polarization with respect to the atomic plane.}
    \label{fig:E_vs_P_scatterplot}
\end{figure}
The spontaneous polarization of the 63 predicted ferroelectrics are summarized in Tab. \ref{tab:in-plane} (in-plane polarization), Tab. \ref{tab:out-of-plane} (out-of-plane polarization) and Tab. \ref{tab:mixed} (both in-plane and out-of-plane polarization).
In Fig. \ref{fig:E_vs_P} we show three examples of the energy as a function of polarization along the adiabatic path obtained by linear interpolation as well as the NEB optimized path. The first one is Ge$_2$S$_2$, which have been subjected to extensive theoretical scrutiny as a prototypical example of a 2D ferroelectric \cite{Fei2016, polarization_2D_monochalcogenides, Wang2017, Barraza-Lopez2021}. In that case the linear interpolation and NEB paths largely coincide, which reflect the simplicity of the ferroelectric transition that involves a small displacement of atoms long the polarization direction. The second example is Li$_2$F$_2$S$_2$, which exhibits a somewhat more complicated path where the different elements undergo displacements in various directions along the path. This is reflected in a large difference between the linearly interpolated path and the NEB path.

\begin{table*}[!htb]
	\begin{center}
		\begin{tabular}{c|c|c|c|c|c|c|c|c}
			Name & Spacegroup & P$_{\parallel}$ $[\text{pC}/\text{m}]$ & $\Delta\text{E}$ $[\text{meV}/\text{Å}^2]$ & $\mathcal{E}_{\text{c}}$ $[\text{V}/\text{nm}]$ & Gap [eV] & $\omega$ $[\text{meV}]$ & EH $[\text{eV}/\text{atom}]$ & ID \\
			\hline
			As$_{4}$O$_{6}$ & P2$_{1}$ & 191 & 42.2 & 45.8 & 3.9 & -36.1 & 0 & 4513280 \\
			Zn$_{2}$As$_{4}$O$_{8}$ & P2$_{1}$ & 313 & 30.7 & 4.51 & 3.6 & 0.00 & 0 & 9001071 \\
			W$_{2}$O$_{12}$Sb$_{4}$ & P2$_{1}$ & 238 & 9.82 & 1.08 & 1.7 & -30.1 & 0.02 & 75595 \\
			Ag$_{2}$Cl$_{2}$Se$_{4}$ & P2$_{1}$ & 359 & 14.4 & - & 1.3 & 0.00 & 0 & - \\
			Ag$_{2}$Cl$_{2}$Te$_{4}$ & P2$_{1}$ & 45.1 & 20.0 & - & 1.3 & -5.33 & 0 & - \\
			Ag$_{2}$Br$_{2}$Se$_{4}$ & P2$_{1}$ & 126 & 13.0 & - & 1.2 & 0.00 & 0 & - \\
			Ag$_{2}$I$_{2}$Te$_{4}$ & P2$_{1}$ & 119 & 19.8 & - & 1.0 & 0.00 & 0.01 & - \\
			Au$_{2}$Cl$_{2}$Te$_{4}$ & P2$_{1}$ & 18.3 & 3.88 & - & 0.98 & 0.00 & 0.1 & - \\
			Au$_{2}$Br$_{2}$Te$_{4}$ & P2$_{1}$ & 25.7 & 5.27 & - & 1.0 & 0.00 & 0.05 & - \\
			Au$_{2}$I$_{2}$Te$_{4}$ & P2$_{1}$ & 60.4 & 8.57 & - & 0.71 & 0.00 & 0.06 & - \\
			Na$_{2}$Nb$_{2}$Cl$_{12}$ & P2$_{1}$ & 32.2 & 2.26 & 1.66 & 2.3 & -3.55 & 0 & 8103950 \\
			Zn$_{2}$Te$_{2}$(N$_{2}$H$_{4}$)$_{2}$ & P2$_{1}$ & 202 & 9.12 & - & 1.8 & 0.00 & 0.1 & 4113794 \\
			K$_{2}$(CHO$_3$)$_2$ & P2$_{1}$ & 44.4 & 0.172 & - & 4.5 & 0.00 & 0 & 9016039 \\
			As$_{2}$O$_{8}$Te$_{2}$(OH)$_{2}$ & P2$_{1}$ & 395 & 2.68 & - & 3.4 & 0.00 & 0 & 425501 \\
			HgH$_{2}$S$_{2}$ & C2 & 323 & 10.8 & - & 2.4 & -13.6 & 0 & - \\
			NiZrF$_{6}$ & C2 & 281 & 31.1 & 18.0 & 0.41 & -10.1 & 0.2 & - \\
			As$_{4}$O$_{6}$ & Pc & 502 & 48.3 & - & 4.3 & -13.8 & 0 & 9014252 \\
			As$_{2}$O$_{6}$Sb$_{2}$ & Pc & 783 & 197 & - & 4.0 & -57.5 & 0 & 9015432 \\
			Sn$_{2}$H$_{2}$O$_{6}$P$_{2}$ & Pc & 301 & 313 & - & 4.3 & -48.9 & 0 & 4328407 \\
			Zn$_{2}$O$_{6}$Se$_{2}$(H$_2$O)$_2$ & Pc & 710 & 166 & - & 4.5 & -52.0 & 0 & 78916 \\
			Nb$_{2}$Cl$_{4}$O$_{2}$ & Pmm2 & 213 & 0.739 & 0.816 & 1.0 & 0.00 & 0 & - \\
			Nb$_{2}$Br$_{4}$O$_{2}$ & Pmm2 & 199 & 0.626 & 0.811 & 1.0 & 0.00 & 0 & 416669 \\
			Nb$_{2}$I$_{4}$O$_{2}$ & Pmm2 & 183 & 5.17 & - & 1.0 & -38.9 & 0 & 36255 \\
			Mo$_{2}$Br$_{4}$O$_{4}$ & Pmc2$_{1}$ & 240 & 11.7 & 2.59 & 1.4 & -44.4 & 0 & 422483 \\
			W$_{2}$Cl$_{4}$O$_{4}$ & Pmc2$_{1}$ & 191 & 5.97 & 1.16 & 2.2 & -31.1 & 0 & 28510 \\
			Sr$_{2}$H$_{8}$O$_{6}$ & Pmc2$_{1}$ & 234 & 45.0 & - & 4.4 & -22.8 & 0 & 15366 \\
			Cu$_{2}$Hg$_{2}$Cl$_{2}$Se$_{2}$ & Pmc2$_{1}$ & 453 & 50.3 & - & 0.83 & -7.91 & 0.009 & 1001109 \\
			Hf$_{2}$Zr$_{2}$S$_{8}$ & Pma2 & 601 & -48.9 & 3.79 & 1.1 & 0.00 & 0.2 & - \\
			Hf$_{2}$Zr$_{2}$Se$_{8}$ & Pma2 & 579 & -36.7 & 5.00 & 0.84 & 0.00 & 0.2 & - \\
			Pb$_{4}$O$_{4}$ & Pca2$_{1}$ & 290 & 10.8 & 0.860 & 2.6 & -20.8 & 0 & 36250 \\
			As$_{4}$S$_{6}$ & Pmn2$_{1}$ & 530 & 19.9 & - & 2.3 & 0.00 & 0.0002 & 9008211 \\
			As$_{4}$Se$_{6}$ & Pmn2$_{1}$ & 494 & 15.7 & 2.69 & 1.7 & 0.00 & 0.002 & 9011471 \\
			Ga$_{2}$In$_{2}$S$_{6}$ & Pmn2$_{1}$ & 922 & 139 & 3.56 & 2.0 & -21.7 & 0 & 62340 \\
			Ga$_{2}$Cl$_{2}$Te$_{2}$ & Pmn2$_{1}$ & 580 & 113 & 5.29 & 2.2 & -18.0 & 0 & 7221395 \\
			Cu$_{2}$C$_{2}$Cl$_{2}$O$_{2}$ & Pmn2$_{1}$ & 252 & 31.6 & 7.85 & 2.5 & -12.1 & 0.09 & 63490 \\
			$\alpha$-Ge$_{2}$S$_{2}$ & Pmn2$_{1}$ & 490 & 30.5 & 1.63 & 1.7 & -25.2 & 0.03 & 2107064 \\
			$\alpha$-Ge$_{2}$Se$_{2}$ & Pmn2$_{1}$ & 357 & 6.66 & 0.482 & 1.1 & -14.4 & 0.02 & 9008783 \\
			$\alpha$-Ge$_{2}$Te$_{2}$ & Pmn2$_{1}$ & 314 & 2.30 & 0.190 & 0.81 & -11.0 & 0.05 & 638005 \\
			$\alpha$-Sn$_{2}$O$_{2}$ & Pmn2$_{1}$ & 305 & 23.4 & 2.09 & 2.5 & -25.4 & 0.05 & 20624 \\
			$\alpha$-Sn$_{2}$S$_{2}$ & Pmn2$_{1}$ & 292 & 3.55 & 0.300 & 1.4 & -13.2 & 0.04 & 1527226 \\
			$\alpha$-Sn$_{2}$Se$_{2}$ & Pmn2$_{1}$ & 219 & 0.960 & 0.109 & 0.92 & -7.29 & 0.04 & 1537675 \\
			$\alpha$-Sn$_{2}$Te$_{2}$ & Pmn2$_{1}$ & 138 & 0.0900 & 0.0147 & 0.60 & -3.64 & 0.06 & 652743 \\
			$\beta$-Ge$_{2}$S$_{2}$ & Pmn2$_{1}$ & 201 & 2.10 & 0.238 & 1.8 & -13.0 & 0.05 & - \\
			$\beta$-Ge$_{2}$Se$_{2}$ & Pmn2$_{1}$ & 161 & 0.589 & 0.0925 & 1.7 & -7.68 & 0.05 & - \\
			$\beta$-Sn$_{2}$S$_{2}$ & Pmn2$_{1}$ & 123 & 0.374 & 0.0750 & 1.8 & -8.07 & 0.06 & - \\
			$\beta$-Sn$_{2}$Se$_{2}$ & Pmn2$_{1}$ & 107 & 0.148 & 0.0345 & 1.6 & -4.93 & 0.07 & - \\
			As$_{2}$Sb$_{2}$ & Pmn2$_{1}$ & 382 & 7.50 & - & 0.75 & -18.3 & 0.09 & - \\
			P$_{2}$Sb$_{2}$ & Pmn2$_{1}$ & 346 & 6.56 & 0.678 & 0.80 & -23.0 & 0.2 & - \\
			Cd$_{2}$O$_{6}$Se$_{2}$(H$_2$O)$_2$ & Pmn2$_{1}$ & 722 & 178 & - & 4.1 & -52.3 & 0 & 2007038
		\end{tabular}
	\end{center}
	\caption{Table of materials having the spontaneous polarization aligned with the atomic plane. The columns show the stoichiometry, space group, magnitude of spontaneous polarization (P$_\parallel$), energy difference between the polar and the non-polar structures ($\Delta E$), coercive field ($\mathcal{E}_\mathrm{c}$), band gap in the polar structure, lowest
    phonon energy of the non-polar structure at the $\Gamma$-point ($\omega$), energy above the convex hull (EH) and finally the ICSD or COD identifier (ID).}
	\label{tab:in-plane}
\end{table*}
\begin{table*}[!htb]
	\begin{center}
		\begin{tabular}{c|c|c|c|c|c|c|c|c}
			Name & Spacegroup & P$_{\perp}$ $[\text{pC}/\text{m}]$ & P$_{\text{top}}$ & $\Delta\text{E}$ $[\text{meV}/\text{Å}^2]$ & $\mathcal{E}_{\text{c}}$ $[\text{V}/\text{nm}]$ & Gap [eV] & EH $[\text{eV}/\text{atom}]$ & ID \\
			\hline
			CuInP$_{2}$S$_{6}$ & P3 & 9.98 & (2/3, 1/3) & 14.7 & 58.2 & 1.5 & 0 & - \\
			CuInP$_{2}$Se$_{6}$ & P3 & 7.34 & (2/3, 1/3) & 8.17 & 13.7 & 0.49 & 0 & 71969 \\
			CuBiP$_{2}$Se$_{6}$ & P3 & 5.02 & (1/3, 2/3) & 8.14 & 27.3 & 1.0 & 0 & 4327329 \\
			AgBiP$_{2}$Se$_{6}$ & P3 & 2.79 & (2/3, 1/3) & 0.336 & 3.84 & 1.1 & 0 & 4327327 \\
			InP & P3m1 & 25.1 & (0, 0) & 3.45 & 1.69 & 1.1 & 0.4 & - \\
			CSiF$_{2}$ & P3m1 & 1.31 & (0, 0) & 295 & 10.7 & 1.9 & 0.6 & - \\
			CSiH$_{2}$ & P3m1 & 38.8 & (0, 0) & 544 & - & 3.8 & 0 & - \\
			In$_{2}$Se$_{3}$ & P3m1 & 33.8 & (1/3, 2/3) & 70.0 & 9.01 & 0.75 & 0.008 & -
		\end{tabular}
	\end{center}
	\caption{Same as Tab. \ref{tab:in-plane}, but for materials with purely out-of-plane polarization (P$_\perp$). In addition we state the topological in-plane polarization (P$_\mathrm{top}$) in units of the 2D $\Gamma$-centered polarization lattice.}
	\label{tab:out-of-plane}
\end{table*}
\begin{table*}[!htb]
	\begin{center}
		\begin{tabular}{c|c|c|c|c|c|c|c|c|c}
			Name & Spacegroup & P$_{\parallel}$ $[\text{pC}/\text{m}]$ & P$_{\perp}$ $[\text{pC}/\text{m}]$ & $\Delta\text{E}$ $[\text{meV}/\text{Å}^2]$ & $\mathcal{E}_{\text{c}}$ $[\text{V}/\text{nm}]$ & Gap [eV] & $\omega$ $[\text{meV}]$ & EH $[\text{eV}/\text{atom}]$ & ID \\
			\hline
			In$_{2}$Te$_{4}$ & P1 & 253 & 5.98 & 0.496 & 0.0935 & 0.49 & 0.00 & 0.2 & 501 \\
			TiZr$_{3}$S$_{8}$ & P1 & 608 & 0.140 & -42.8 & - & 1.0 & 0.00 & 0.2 & - \\
			HfZr$_{3}$S$_{8}$ & P1 & 600 & 0.411 & -46.9 & - & 1.2 & 0.00 & 0.2 & - \\
			Li$_{2}$F$_{2}$S$_{2}$ & P1 & 114 & 1.70 & 1.87 & 1.31 & 1.7 & -2.13 & 0.5 & - \\
			Hf$_{3}$ZrS$_{8}$ & Pm & 602 & 0.230 & -51.2 & - & 1.1 & 0.00 & 0.2 & - \\
			Hf$_{3}$ZrSe$_{8}$ & Pm & 580 & 0.261 & -38.7 & 3.66 & 0.84 & 0.00 & 0.2 & -
		\end{tabular}
	\end{center}
	\caption{Same as Tab. \ref{tab:in-plane}, but for materials with in-plane polarization (P$_\parallel$) as well as out-of-plane polarization (P$_\perp$).}
	\label{tab:mixed}
\end{table*}
Finally, the case In$_2$Te$_4$ has in-plane as well as out-of-plane components of the polarization. Here the switching barrier is lowered by an order of magnitude compared to the linear interpolation. The centrosymmetric state is seen to reside in a local minimum and is thus metastable, which could have important implications for the thermodynamical properties. In fact, for these three examples it is only the case of Ge$_2$S$_2$ that appears to be representable by the Landau free energy type of expression $F=\alpha P^2+\beta P^4$ with $\alpha<0$ relevant for second order phase transitions. When the centrosymmetric phase is metastable (like  In$_2$Te$_4$) the path may be (roughly) fitted to a sixth order polynomial of the form $F=\alpha P^2+\beta P^4+\gamma P^6$ with $\alpha>0$ and $\beta<0$ and the local minimum might indicate a first order phase transition to the ferroelectric phase if the free energy of the metastable state is lowered more rapidly than the polar minimum with increasing temperature. This could be estimated by comparing the entropy originating from phonons in the two minima, but we will leave a detailed analysis of the thermodynamics to future work. 

We have divided all the ferroelectrics into two classes depending on whether the non-polar state is metastable or not. As a sanity check we have calculated the phonon spectrum in the non-polar phase for all materials and compared with the NEB curvature. The results are shown in Fig. \ref{fig:Phonon_scatterplot} and it is confirmed that materials with a positive NEB curvature in the non-polar phase are characterized by a lack of imaginary phonon frequencies (represented as negative phonon energies in Fig. \ref{fig:Phonon_scatterplot}). We find a total of 16 materials with metastable centrosymmetric phases. For the materials that are unstable in the non-polar phase we have analyzed the symmetry of the lowest imaginary phonon mode. In all cases we find the mode to be polar and all the materials in this work can thus be regarded as proper ferroelectrics.

Next, we distinguish between materials having in-plane polarization, out-of-plane polarization and materials where the polarization have both in-plane and out-of-plane components. In Fig. \ref{fig:E_vs_P_scatterplot} we present a scatter plot of the polarization versus energy difference between the polar and centrosymmetric states color coded according to the direction of polarization with respect to the atomic plane. In Tabs. \ref{tab:in-plane}-\ref{tab:mixed} we list all of the materials including spacegroups, magnitude of spontaneous polarization, band gap, coercive field and energy above convex hull. For the materials that are known in bulk form as layered van der Waals bonded structures we also state the identifier of the bulk material from either ICSD \cite{Allmann2007} (ID $<10^6$) or COD (ID $>10^6$).  We note that we find a few materials with a negative energy difference signifying that the centrosymmetric phase is more stable. These are Hf/Zr alloys of transition metal dichalcogenides and and their Janus structures. The present calculations simply show that the (centrosymmetric) T-phase is more stable than the polar H-phase, which is the one appearing in the C2DB.

In Tab. \ref{tab:in-plane} we display the 2D ferroelectrics with an in-plane polar axis. Most of the materials are situated within 0.1 eV per atom of the convex hull and can thus be regarded as being thermodynamically stable (with the exceptions of Hf$_2$Zr$_2$S$_8$ and Hf$_2$Zr$_2$Se$_8$ discussed above). It should also be noted that the far majority of the structures are derived from layered bulk materials (implied by the presence of the ID reference) and are thus expected to be exfoliable from known bulk materials. The spontaneous 2D polarization is generally on the order of a few hundred pC/m. The largest polarization is found for Ga$_2$In$_2$S$_6$ with a value of 922 pC/m, but it also exhibits a rather high barrier of 139 $\text{meV}/\text{\AA}^2$, which could imply that it is not switchable under realistic conditions. We emphasize again that in the present work we have defined materials to be ferroelectric if an adiabatic path exists that connects the structure to the closest centrosymmetric phase. There may be several other pyroelectric materials in the C2DB where an adiabatic path to a non-polar phase exists that we have not found in the present study. In addition, for a given material identified as ferroelectric there may exist a different path that lowers the barrier by circumventing the centrosymmetric structure. This has, for example been found for the case of Ge$_2$Se$_2$ \cite{Wang2017} and the stated values for barriers and coercive fields in this work should be regarded as upper limits. This does, not, however, affect the reported value of the spontaneous polarization as discussed above. Compounds in Tab. \ref{tab:in-plane} with low barriers are of particular interest, since these are expected to be easily switchable. Besides the Ge and Sn chalcogenides, which have already been discussed extensively in literature \cite{SnTe_experiment, SnS_experiment, SnSe_experiment, SnSedomainswitching,polarization_2D_monochalcogenides, XTes_paper,Wang2017,DW_DFT_Coercive2022}, we find C$_2$H$_2$K$_2$O$_6$, Nb$_2$Na$_2$Cl$_{12}$ and As$_2$H$_2$Te$_2$O$_{10}$ as interesting candidates that are easily switchable. In addition, the niobium oxide dihalides Nb$_2$O$_2$X$_4$ (X=Cl,Br,I) have low switching barriers and large spontaneous polarization. These have, however, been shown to exhibit an anti-ferroelectric ground state \cite{Niobiumoxides}, which has not been considered here. While the barrier itself does not yield a quantitative estimate of the ease with which a ferroelectric may switch polarisation state it is expected to constitute a rough qualitative estimate. For a more precise estimate one may calculate the coercive field for coherent monodomain switching. For the materials where we succeeded in converging the NEB calculation we have calculated this field according to Eq. \eqref{eq:field}.

From Tab. \ref{tab:in-plane} it is clear that the calculated coercive fields largely correlate with the barrier and the lowest fields are found for materials with small barriers. We have not put additional effort into converging all NEB calculations since the actual switching mechanism in ferroelectrics typically involve migration of domain walls rather than coherent monodomain switching and the estimated coercive fields may thus be orders of magnitude larger than experimentally relevant coercive fields \cite{DW_DFT_Coercive2022}. Nevertheless, for the Ge and Sn monochalcogenides the fields required for switching the polarization state through domain wall migration have been shown to be largely correlated with the energy barrier for switching \cite{DW_DFT_Coercive2022} and the {\it relative} values of coercive fields in Tab. \ref{tab:in-plane} may be regarded as a rough measure of the hardness with which the materials can be switched by an external electric field.

For the out-of-plane ferroelectrics stated in Tab. \ref{tab:out-of-plane} it is clear from direct inspection there exist simple paths that do not pass through inversion symmetric structures. Referring to the example of AgBiP$_2$Se$_6$ in Fig. \ref{fig:materialprototypes} it is easy to see that there is a path connecting states of opposite polarity without passing through an inversion symmetric point. This involves shifting the Ag atom from the top Se layer to the bottom Se layer. The path passes through a structure with the non-polar point group 32, which does not have inversion symmetry. For most of the materials there exist another adiabatic path passing through a inversion symmetric point, but the energy barrier along the 32 path is much lower. Our general workflow only identifies paths involving centrosymmetric point, but we have recalculated the barriers along the simple paths (not passing through centrosymmetric structures) for all the materials in Tab. \ref{tab:out-of-plane}.
\begin{figure*}[!htb]
    \centering
    \includegraphics[width=1.9\columnwidth]{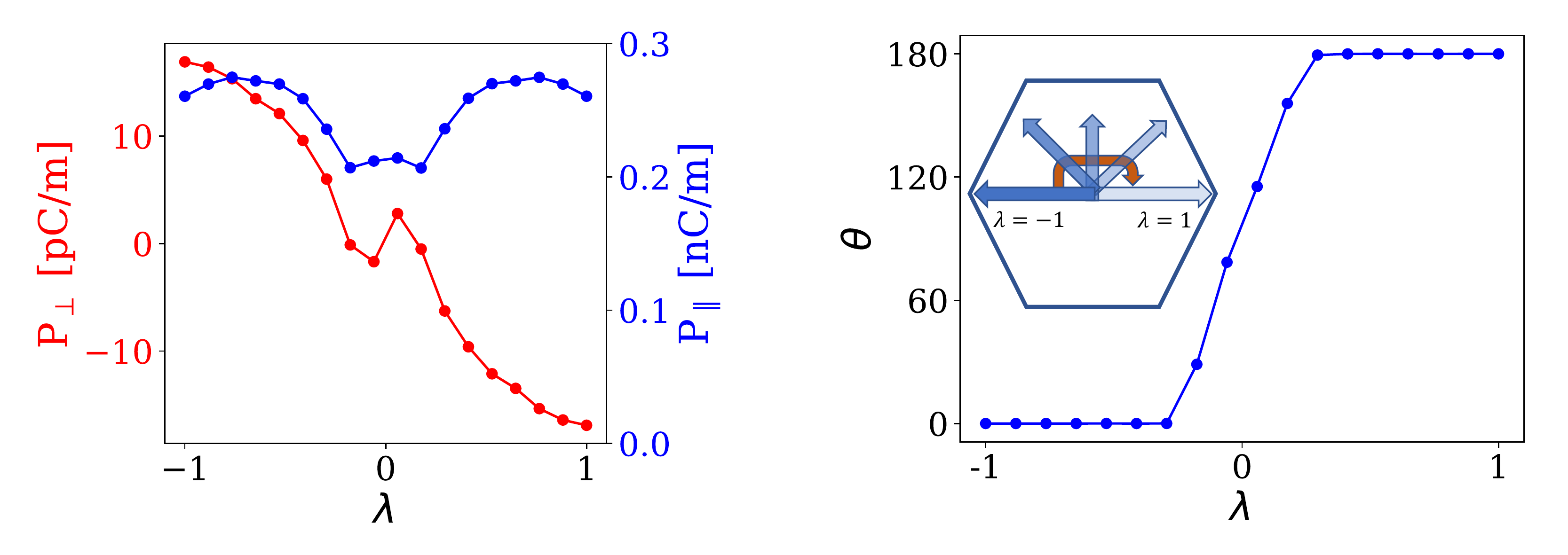}
    \caption{Polarization switching of In$_2$Se$_3$. Left: the out-of-plane (P$_\perp$) and magnitude of in-plane (P$_\parallel$) polarization along a path that switches between two inversion related high symmetry points in the Wigner-Seitz cell. Right: the angle of in-plane polarization along the switching path.}
    \label{fig:P_InSe}
\end{figure*}

In general, we find that the magnitude of polarization is significantly smaller in materials with purely out-of-plane polarization compared to those with in-plane polarization. The experimentally known structures ABP$_2$Se$_6$ (A=Ag, Cu and B=In, Bi) in Tab. \ref{tab:out-of-plane} have a small out-of-plane dipole originating from the A atom being displaced from the center of the layer towards the top or bottom (see. Fig. \ref{fig:materialprototypes}) and the remaining materials are buckled honeycomb structures with inequivalent atoms that naturally gives rise to a small dipole. The polar state in these materials involves a significant displacement of the atoms, but the magnitudes of polarization are rather small due small values of the ($\sim0.1e$) $z$-components of Born effective charge tensors \cite{Gjerding2021} for the displaced atoms. 

All of the materials are classified as purely out-of-plane ferroelectrics due to a three-fold rotational symmetry in the plane. However, since the formal polarization Eq. \eqref{eq:formal_polarization} is not single valued, the in-plane polarization is allowed to be non-vanishing. In particular, for the case of three-fold rotational symmetry the allowed values are $(0, 0)$, $(1/3, 2/3)$ and $(2/3, 1/3)$ in units of cell vectors divided by cell area (for a hexagonal cell with $120^\circ$ angle between vectors). Such a "topological polarization" is not switchable, but enforces gapless states at any zigzag terminated ribbon or nanoflake \cite{Gibertini2015, Benalcazar2019, Sodequist2022}. Interestingly, all of the experimentally known materials in Tab. \ref{tab:out-of-plane} exhibit non-trivial topological polarization. 

The case of In$_2$Se$_3$ deserves a brief discussion in terms of topological properties. In contrast to the remaining materials in Tab. \ref{tab:out-of-plane}, this material breaks the three-fold rotational symmetry along the adiabatic path. The allows for a coupling of the out-of-plane polarization to the in-plane polarization, since the in-plane polarization is only protected by topology in the relaxed polar structures. In Fig. \ref{fig:P_InSe} we show the in-plane and out-of-plane polarization of In$_2$Se$_3$ along the adiabatic path and indeed observe that both components of polarization are switched simultaneously. While the out-of-plane components vanishes halfway along the path, the magnitude of the in-plane component only varies marginally, and simply rotates from one high symmetry point to another. It thus changes topology from (1/3, 2/3) to (2/3, 1/3) along the path. We note that the coupling of polarization components has been observed experimentally in thin films of In$_2$Se$_3$, \cite{alphaIn2Se3experiment} but to our knowledge the topological properties of the polarization has not been unravelled prior to the present work.

In Tab. \ref{tab:mixed} we list all materials that have in-plane as well as out-of-plane components of the spontaneous polarization. The only point groups that allow for this are $1$ and $m$. In the latter case the mirror plane has to be orthogonal to the atomic plane to allow for a polarization that is not purely in-plane. Again, we find a few alloys of transition metal dichalcogenides in the H-phase that are found to be more stable in the T-phase (as implied by negative values of $\Delta E$). These are probably not switchable under realistic conditions and it is dubious if it is even possible to synthesize any of them. Instead we wish to highlight the case of In$_2$Te$_4$, which is the only material with mixed polarization that is experimentally known as a van der Waals bonded bulk material and thus may be easy to exfoliate. The out-of-plane component of polarization is much smaller than the in-plane component, but since the components are coupled along the switching path the in-plane polarization can be switched by application of a purely out-of-plane field. This could have significant practical consequences for the operation of ferroelectric devices based on 2D materials since a large out-of-plane field can be implemented by rather simple means using top and bottom gates. 
\begin{table*}[!htb]
	\begin{center}
		\begin{tabular}{c|c|c|c|c|c|c|c}
			Name & Spacegroup & Gap [eV] & J [meV] & Spin [$\hbar$] & Axis & EH $[\text{eV}/\text{atom}]$ & ID \\
			\hline
			ReAu$_{2}$F$_{6}$ & P1 & 0.18 & - & 1/2 & 3D & 0.2 & - \\
			NiPS$_{3}$ & P1 & 0.075 & 5.9 & 1/2 & 3D & 0.3 & 646140 \\
			Cu$_{2}$I$_{4}$O$_{1}$$_{2}$ & P2$_{1}$ & 0.73 & 3.3 & 1/2 & $\parallel$ & 0.03 & 4327 \\
			Cr$_{2}$Cu$_{2}$P$_{4}$S$_{1}$$_{2}$ & P2$_{1}$ & 1.1 & 1.2 & 3/2 & $\parallel$ & 0 & 1000355 \\
			CoZrBr$_{6}$ & C2 & 0.39 & - & 1/2 & $\parallel$ & 0 & - \\
			VAgP$_{2}$Se$_{6}$ & C2 & 0.35 & 1.9 & 1 & $\parallel$ & 0.004 & 1509506 \\
			Mn$_{2}$H$_{4}$O$_{8}$S$_{2}$ & Pc & 2.1 & -2.1 & 2 & $\parallel$ & 0.08 & 74810 \\
			VF$_{2}$O & Pmm2 & 0.76 & 13 & 1/2 & $\parallel$ & 0.001 & - \\
			VCl$_{2}$O & Pmm2 & 0.81 & -26 & 1/2 & $\parallel$ & 0 & 24380 \\
			VBr$_{2}$O & Pmm2 & 0.77 & -14 & 1/2 & $\parallel$ & 0 & 24381 \\
			VI$_{2}$O & Pmm2 & 0.50 & 6.8 & 1/2 & $\parallel$ & 0 & - \\
			NiC$_{6}$Cl$_{2}$H$_{4}$N$_{2}$ & Pmm2 & 0.86 & 9.1 & 1 & $\parallel$ & 0.2 & 7227895 \\
			Mn$_{2}$Cl$_{2}$Sb$_{2}$S$_{4}$ & Pmc2$_{1}$ & 0.35 & -3.0 & 2 & $\parallel$ & 0.08 & 151925 \\
			Mn$_{2}$Br$_{2}$Sb$_{2}$Se$_{4}$ & Pmc2$_{1}$ & 0.26 & -3.2 & 2 & $\parallel$ & 0.09 & 1528451 \\
			Ti$_{4}$Cl$_{4}$Se$_{4}$ & Pmn2$_{1}$ & 0.042 & - & 1/2 & $\parallel$ & 0.1 & - \\
			Ti$_{4}$Br$_{4}$Se$_{4}$ & Pmn2$_{1}$ & 0.063 & - & 1/2 & $\parallel$ & 0.1 & - \\
			VClBr & P3m1 & 1.3 & -4.1 & 3/2 & $\perp$ & 0.01 & - \\
			VClI & P3m1 & 1.1 & -2.5 & 3/2 & $\perp$ & 0.06 & - \\
			VBrI & P3m1 & 1.2 & -1.5 & 3/2 & $\perp$ & 0.02 & - \\
			VSSe & P3m1 & 0.013 & 58 & 1/2 & $\perp$ & 0 & - \\
			VSeTe & P3m1 & 0.12 & 74 & 1/2 & $\perp$ & 0.01 & - \\
			Nb$_{3}$Cl$_{8}$ & P3m1 & 0.21 & - & 1/6 & $\perp$ & 0 & 408645 \\
			Nb$_{3}$Br$_{8}$ & P3m1 & 0.29 & - & 1/6 & $\perp$ & 0 & 1539108 \\
			Nb$_{3}$I$_{8}$ & P3m1 & 0.19 & - & 1/6 & $\perp$ & 0 & 1539109 \\
			CrHO$_{2}$ & P3m1 & 0.092 & 3.0 & 3/2 & $\perp$ & 0.2 & 9012135 \\
			Cr$_{2}$P$_{2}$Se$_{6}$ & P31m & 0.43 & 11 & 3/2 & $\perp$ & 0.01 & 626521
		\end{tabular}
	\end{center}
	\caption{Dynamically stable 2D polar magnetic materials that exhibits a band gap. In addition to the columns shown in Fig. \ref{tab:in-plane}, we also state the nearest neighbor exchange constant (positive for ferromagnets) and the spin carried by the magnetic atoms. The polar axis follows from the point group symmetries and is stated for convenience.}
	\label{tab:magnets}
\end{table*}
\section{Magnetic polar materials}\label{sec:mag}
The magnetic polar materials deserve a special treatment for several reasons. First of all, the interest in 2D magnetism has exploded in recent years and novel 2D magnets are interesting in their own right - polar or not. Since polar materials are pyroelectric by definition all the polar magnetic materials from C2DB can be regarded as being multiferroic and these compounds may exhibit coupling between magnetic order and polar order. Moreover, polar magnetic materials are magnetoelectric, but we will leave the study of the magnetoelectric effect in 2D to future work, since it would be inconsistent to exclude non-polar non-centrosymmetric compounds in a systematic high throughput screening of such magnetoelectrics. In Tab. \ref{tab:magnets} we list all dynamically stable 2D magnetic materials that are polar and have a finite band gap. In addition to the gap and energy above the convex hull, we also state the spin state of the magnetic transition metal atoms and the nearest neighbor exchange coupling obtained from collinear energy mapping \cite{Torelli2020a,Torelli2020}. The sign of the nearest neighbor exchange coupling (positive for ferromagnetic interaction) in Tab. \ref{tab:magnets} indicates whether the true ground state is expected to have ferromagnetic or anti-ferromagnetic order. 

A few of the materials in Tab. \ref{tab:magnets} have already been discussed prior to this work. For example, the vanadium oxy-halides VOX$_2$ (X=F, Cl, Br, I) have been shown to be switchable multiferroics \cite{VOX2_multiferroics, VOF2paper}. VOCl$_{2}$ and VOBr$_{2}$ exhibit anti-ferromagnetic order in the gapped ground state while the ferromagnetic state is metallic. As a consequence, they are not captured by our workflow since all materials in C2DB are reported in their ferromagnetic state. Furthermore, it has been argued that it is necessary to apply beyonf-PBE approaches such as HSE or PBE+U in order to calculated the band gap accurately for these materials. VOF$_{2}$ and VOI$_{2}$ may thus be predicted to be gapped and switchable in the ferromagnetic state if a more accurate functional is applied \cite{VOX2_multiferroics}. 

None of the materials in Tab. \ref{tab:magnets} have an adiabatic switching path that passes through a centrosymmetric point why they are not present in Tabs. \ref{tab:in-plane}-\ref{tab:mixed}. This does not, however, imply that none of the compounds are ferroelectric and due to the intriguing possibility of having coexistence of ferromagnetic and ferroelectric orders it is worthwhile to investigate whether any of these may be switched through an alternative path. Here we will just focus on two examples - VAgP$_2$Se$_6$ and Cr$_2$P$_2$Se$_6$ - and leave a systematic study of 2D multiferroics to future work. 

\begin{figure*}[!htb]
    \centering
    \includegraphics[width=2\columnwidth]{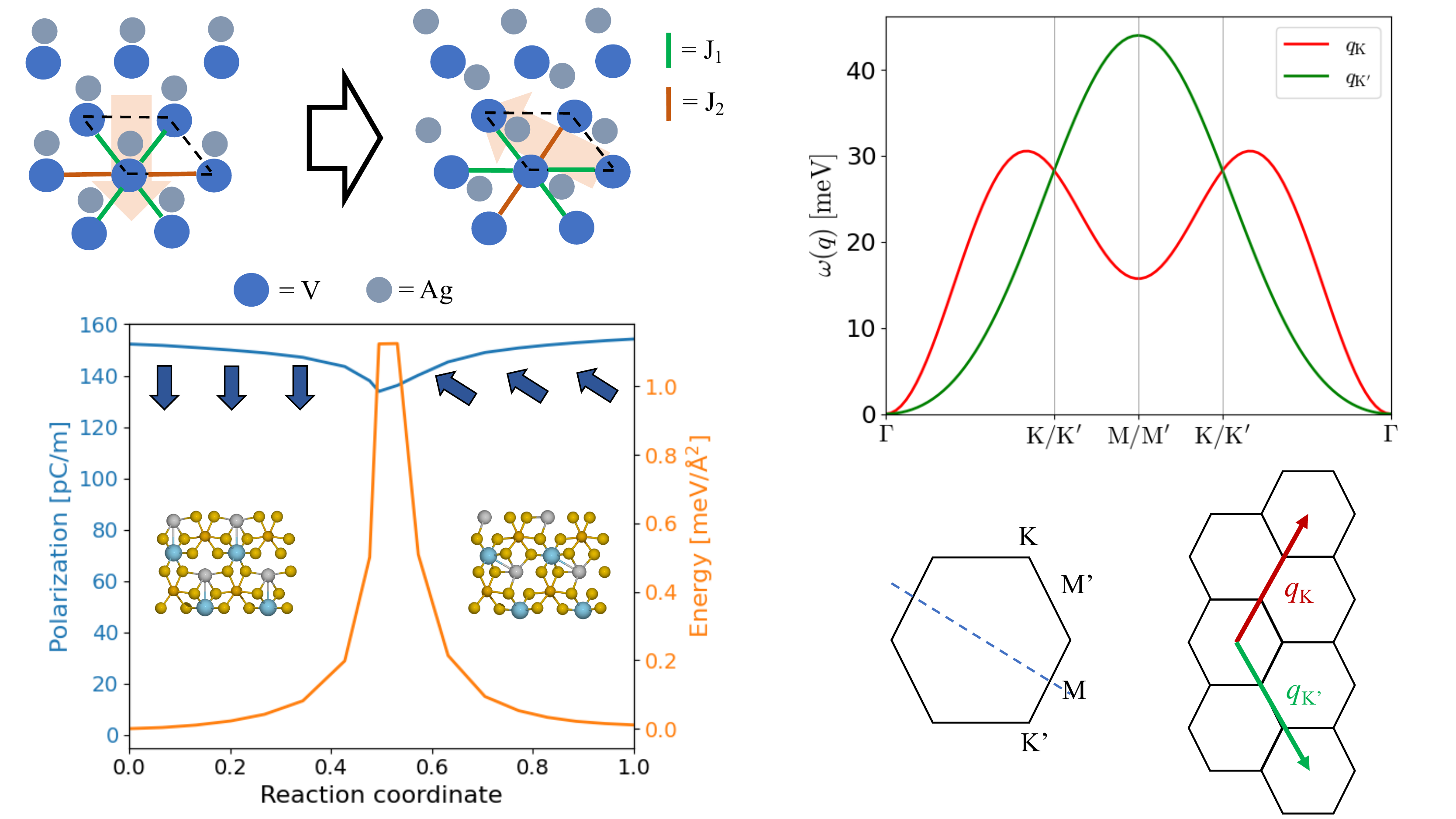}
    \caption{Polar states and magnetic excitations of VAgP$_2$Se$_6$. Left: Polarization and energy along the adiabatic path that switches the polarization by 120$^\circ$. The top shows a schematic view of the Ag/V lattice with the two inequivalent nearest neighbor exchange interactions indicated in the two polarization states. Right: Magnon energies along different directions in the Brillouin zone (polar axis indicated by dashed line) derived from nearest neighbor interactions. We note that $\mathrm{K}$ and $\mathrm{K}'$ are equivalent but are given different labels here to distinguish the $\Gamma \mathrm{K}$ and $\Gamma\mathrm{K}'$ directions.}
    \label{fig:AgVP2Se6}
\end{figure*}

Bulk VAgP$_2$Se$_6$ was synthesized and characterized in Ref. \cite{OUVRARD19881199}, where it was found to have space group C2 and to order ferromagnetically below 18 K. The vanadium atoms are in the oxidation state V$^{3+}$, which implies a $d^2$ configuration with a magnetic moment of 2 $\mu_\mathrm{B}$. The basic structure of the individual layers can be envisioned as a distorted honeycomb lattice of alternating Ag and V atoms as shown in Fig. \ref{fig:AgVP2Se6}. Each V atom thus have one nearest neighbor Ag atom (distance 3.21 $\mathrm{\AA}$) and two next nearest neighbors (distance 3.96 $\mathrm{\AA}$), which produces a polar axis along the short bond. Since each V atom may form three equivalent short bonds the compound has three polarization states where the polarization vectors are rotated by $120^\circ$. The PBE relaxed monolayer in C2DB is dynamically stable and has the same space group as the bulk material. In Fig. \ref{fig:AgVP2Se6} we show the switching path obtained from a NEB calculations between two such rotated polarization states. The barrier is rather low with a value of 1.2 meV/$\mathrm{\AA}^2$ and the structure is thus expected to be easily switchable by application of an external electric field. We also calculate the spontaneous polarization along the path and find that the magnitude is nearly constant, but with an abrupt change in direction at the point where the Ag atom is exactly between the two V atoms. The material thus comprises an example of a system with a switchable discrete three-state polarization, which may be of relevance for non-volatile memory applications. The V atoms constitute the magnetic lattice, which is hexagonal and each V atom thus have six nearest neighbors. These are, however not equivalent due to the polarization and the two exchange constants ($J_1$) orthogonal to the polar axis are distinct from the four exchange constants ($J_2$) along bonds that are at a $30^\circ$ angles to the polar axis as illustrated in Fig. \ref{fig:AgVP2Se6}. We have calculated the two exchange constants by an energy mapping approach \cite{Xiang2013,Torelli2019} using three magnetic configurations in a $2\times2$ repetition of the unit cell. One ferromagnetic state and two states with ferromagnetic chains that are anti-ferromagnetically aligned with neighboring chains and at different angle with respect to the polarization direction. The three energies can be mapped to the nearest neighbor isotropic Heisenberg model
\begin{align}
    H=-\frac{1}{2}J_1\sum_{\langle ij\rangle_\angle}\mathbf{S}_i\cdot\mathbf{S}_j-\frac{1}{2}J_2\sum_{\langle ij\rangle_\perp}\mathbf{S}_i\cdot\mathbf{S}_j,
\end{align}
where $\mathbf{S}_i$ denotes the spin of vanadium site $i$, $\langle ij\rangle_\perp$ denotes sum over nearest neighbors perpendicular to polarization and $\langle ij\rangle_\angle$ denotes sum over the remaining nearest neighbors. We then find that $J_1=5.5$ meV and $J_2=-1.6$ meV. It should be noted that the nearest neighbor exchange coupling stated in Tab. \ref{tab:magnets} differs from $J_1$ calculated here, since the former value was obtained from simple high throughput calculations where only two magnetic configurations were considered. It is rather interesting that the exchange path perpendicular to the polarization direction is anti-ferromagnetic whereas the remaining paths are ferromagnetic. These exchange couplings yield a ferromagnetic ground state, but the sign of $J_2$ introduces magnetic frustration in the system. The polar axis also introduces a small in-plane magnetic anisotropy (roughly 0.06 meV per formula unit), but this will not be easily observable since we predict (a weak) out-of-plane easy-axis in this material. The ground state magnetization is therefore not expected to change when the polarization direction is switched between the three discrete states. Nevertheless, the polarization will be observable from the magnetic excitations, which will have a dispersion that is strongly influenced by the direction of the polar axis. In Fig. \ref{fig:AgVP2Se6} we show the magnon dispersion calculated from linear spin-wave theory \cite{Yosida1996} along two different paths that would be equivalent without the polar axis. Due to the large difference between $J_1$ and $J_2$ one observes a magnon energy at the Brillouin zone boundary (the M-point) that is more than two times smaller in the direction parallel to the polarization compared to the directions that are not parallel (the M'-point). Such a switchable magnons dispersion could perhaps find applications in information processing based on magnonics \cite{Pirro2021}. 

The second case of Cr$_2$P$_2$Se$_6$ has spacegroup P31m, which implies a polar axis out-of-plane. The material exhibits a rather large ferromagnetic nearest neighbor exchange constant of 11 meV and weak out-of-plane easy axis. The structure is similar to the P3 materials of Tab. \ref{tab:out-of-plane}, but the Cr atoms are centered in the atomic plane and the P atoms are shifted slightly to the top or bottom of the layer. Again, the switching path involves a simple shift of the P atoms in the direction orthogonal to the plane, which does not pass through a centrosymmetric point. We have calculated the NEB energy along the path connecting the two states of opposite polarization and find an energy barrier of 8.24 meV/Å$^{2}$. However, the material has a rather small gap of 0.43 eV, which closes along the path and  Cr$_2$P$_2$Se$_6$ is therefore not switchable according to the criterion of an adiabatic path. Nevertheless, the out-of-plane polarization is easily calculated by integrating the dipole density over the $z$-direction, which yields P$_\perp$ = 5.4 pC/m. Although the gap closes along the path, it is highly likely that one may still switch the state of polarization by an external electric field, since the metallic states confined in the 2D layer will not be able to fully screen a transverse field.

We finally mention, that all the present calculations of magnetic materials are collinear and cannot capture potential type-II multiferroics where non-collinear magnetic order introduces a polar axis. This has, for example been found in NiI$_2$ \cite{Song2022} and Hf$_2$VC$_2$F$_2$ \cite{Jun-Jie2018}.
\section{Conclusion and outlook}\label{sec:out}
The starting point of the present work has been the 252 polar insulators in the C2DB. For pyroelectric applications all of these are interesting in their own right and may easily be filtered from the C2DB by the point group alone. The main interest here has been the ferroelectric properties and we have thus calculated the spontaneous polarization of all materials where an adiabatic path that pass through a centrosymmetric structure in close proximity to the polar phase could be identified. These materials are expected to be switchable by an external electric field and we found 49 materials with purely in-plane polarization, 8 materials with purely out-of-plane polarization and 6 materials with mixed components of polarization. In particular, we have recovered most of the well-known 2D ferroelectrics, but have extended the list by a large number of new 2D ferroelectrics. We emphasize that most of the discovered compounds are known as bulk van der Waals bonded materials and are expected to be easily exfoliable. Each of the materials in Tabs. \ref{tab:in-plane}-\ref{tab:mixed} thus warrants a more detailed study of ferroelectric properties with respect to stability under ambient conditions, domain wall formation Curie temperatures etc, but this is out of scope for the present work.

It is worth stressing that the present study is by no means exhaustive. The requirement of having a centrosymmetric non-polar phase that is adiabatically connected to the polar structure is not in general fulfilled for all ferroelectrics. The case of CuInP$_2$Se$_6$ comprises an example where the polar structure has point group 3 and the non-polar structure has the non-centrosymmetric point group 32. It is possible to filter materials where a non-centrosymmetric non-polar structure may exist in a systematic way. For example, if the polar structure contains a two-fold axis and an orthogonal mirror plane (this is the case of half the materials in Tab. \ref{tab:in-plane}) it is expected that the non-polar state (roughly) conserves the mirror and the non-polar state thus cannot reside in a chiral point group, which leaves only five possible non-centrosymmetric points groups that will typically be inconsistent with the Bravais lattice. We have not attempted to perform such analysis here, but merely point at possible future directions for a more systematic search for 2D ferroelectrics.

Finally, we have allocated special attention to the magnetic polar materials present in the C2DB. While none of these are switchable through an adiabatic path passing through a centrosymmetric point, the materials are inherently multiferroic even if they are not ferroelectric and are expected to exhibit non-vanishing Dzyaloshinskii-Moriya interactions, magnetoelectric coupling and a range of other properties that are dependent on broken time-reversal and broken inversion symmetry. We have highlighted the case of VAgP$_2$Se$_6$ (point group 2) where the non-polar structure is in close proximity to a structure of 32 symmetry, which lacks inverson symmetry. It is, however, straightforward to identify three equivalent polar structures and calculate the spontaneous polarization as well as the barrier for switching. We found that the structural space for switching comprises nearly discrete states of polarization, which situates the material rather far from a Landau free energy type of description (at low temperatures). Moreover, we showed that the polar axis introduces highly anisotropic nearest neighbor exchange interactions, which introduces a strong coupling of the magnetic excitations spectrum to the polar state of the material. Again, we believe that all of the compounds listed in Tab. \ref{tab:magnets} warrant a more detailed study and there could be several additional materials that allows for an adiabatic switching path that does not involve a centrosymmetric point.
\bibliography{references}
\end{document}